\documentclass{article}
\usepackage{epsf,epsfig,pstricks,pst-node}
\title{Complexity of Networks}
\author{Russell K. Standish\\
Mathematics, University of New South Wales}

\def\citeyear(#1)#2{\cite{#2}}

\newcommand{\EcoLab}{{\sffamily\slshape
    \mbox{\raisebox{.5ex}{Eco}\hspace{-.4em}{\makebox[.5em]{L}ab}}}}

\begin{document}
\maketitle
\begin{abstract}
Network or graph structures are ubiquitous in the study of complex
systems. Often, we are interested in complexity trends of these system
as it evolves under some dynamic. An example might be looking at the
complexity of a food web as species enter an ecosystem via migration
or speciation, and leave via extinction.

In this paper, a complexity measure of networks is proposed based on
the {\em complexity is information content} paradigm. To
apply this paradigm to any object, one must fix two things: a representation
language, in which strings of symbols from some alphabet describe, or
stand for the objects being considered; and a means of determining
when two such descriptions refer to the same object. With these two
things set, the information content of an object can be computed in
principle from the number of equivalent descriptions describing a
particular object.

I propose a simple representation language for undirected graphs that
can be encoded as a bitstring, and equivalence is a topological
equivalence. I also present an algorithm for computing the complexity of
an arbitrary undirected network.
\end{abstract}

\section{Introduction}

In \cite{Standish01a}, I argue that {\em information
  content} provides an overarching complexity measure that connects
the many and various complexity measures proposed (see 
\cite{Edmonds99} for a review). The idea is fairly simple.
In most cases, there is an obvious {\em prefix-free} representation language
within which descriptions of the objects of interest can be encoded.
There is also a classifier of descriptions that can determine if two
descriptions correspond to the same object. This classifier is
commonly called the {\em observer}, denoted $O(x)$.

To compute the complexity of some object $x$, count the number of
equivalent descriptions $\omega(\ell,x)=$ of length $\ell$ that map to
the object $x$ under the agreed classifier. Then the complexity of $x$
is given in the limit as $\ell\rightarrow\infty$:
\begin{equation}\label{complexity}
{\cal C}(x) = \lim_{\ell\rightarrow\infty} \ell\log N - \log\omega(\ell,x)
\end{equation}
where $N$ is the size of the alphabet used for the representation language.

Because the representation language is prefix-free, every description
$y$ in that language has a unique prefix of length $s(y)$.
The classifier does not care what symbols appear after this unique
prefix. Hence $\omega(\ell,O(y))\geq N^{\ell-s(y)}$, $0\leq C(O(y)) \leq
s(y)$ and so equation (\ref{complexity}) converges.

The relationship of this algorithmic complexity measure to more
familiar measures such as Kolmogorov (KCS) complexity, is given by the coding
theorem\cite[Thm 4.3.3]{Li-Vitanyi97}. Equation (\ref{complexity})
corresponds to the logarithm of the {\em universal a priori
  probability}. The difference between these measures is bounded by a
constant independent of the complexity of $x$.

Many measures of network properties have been proposed, starting with
node count and connectivity (no. of links), and passing in no         
particular order through cyclomatic number (no. of independent loops),
spanning height (or width), no. of spanning trees, distribution of
links per node and so on. Graphs tend to be classified using these
measures --- {\em small world} graphs tend to have small spanning
height relative to the number of nodes and {\em scale free} networks
exhibit a power law distribution of node link count.

Some of these measures are related to graph complexity, for example
node count and connectivity can be argued to be lower and upper bounds          
of the network complexity respectively. However, none of the proposed
measures gives a theoretically satisfactory complexity measure, which
in any case is context dependent (ie dependent on the {\em observer}
$O$, and the representation language). 

In this paper we shall consider only undirected graphs, however the
extension of this work to directed graphs should not pose too great a
problem. In setting the classifier function, we assume that only the
graph's topology counts --- positions, and labels of nodes and links
are not considered important. Clearly, this is not appropriate for all
applications, for instance in food web theory, the interaction
strengths (and signs) labeling each link is crucially important.

The issue of representation language, however is far more problematic.
In some cases, eg with genetic regulatory networks, there may be a
clear representation language, but for many cases there is no uniquely
identifiable language. However, the {\em invariance theorem}\cite[Thm
2.1.1]{Li-Vitanyi97} states that the difference in complexity
determined by two different {\em Turing complete} representation
languages (each of which is determined by a universal Turing machine)
is at most a constant, independent of the objects being measured.
Thus, in some sense it does not matter what representation one picks
--- one is free to pick a representation that is convenient, however
one must take care with non Turing complete representations.

In the next section, I will present a concrete graph description
language that can be represented as binary strings, and is amenable to
analysis. The quantity $\omega$ in eq (\ref{complexity}) can be simply
computed from the size of the automorphism group, for which
computationally feasible algorithms exist\cite{McKay81}.

The notion of complexity presented in this paper naturally marries
with thermodynamic entropy $S$\cite{Layzer88}:
\begin{equation}\label{Layzer}
S_{\max}={\cal C} + S
\end{equation}
where $S_{\max}$ is called {\em potential entropy}, ie the largest
possible value that entropy can assume under the specified
conditions. The interest here is that a dynamical process updating
network links can be viewed as a dissipative system, with links being
made and broken corresponding to a thermodynamic flux. It would be
interesting to see if such processes behave according the maximum
entropy production principle\cite{Dewar03} or the minimum entropy
production principle\cite{Prigogine80}.

In artificial life, the issue of complexity trend in evolution is
extremely important\cite{Bedau-etal00}. I have explored the complexity
of individual Tierran organisms\cite{Standish03a,Standish04c}, which,
if anything, shows a trend to simpler organisms. However, it is
entirely plausible that complexity growth takes place in the network
of ecological interactions between individuals. For example, in the
evolution of the eukaryotic cell, mitochondria are simpler entities
than the free-living bacteria they were supposedly descended. A
computationally feasible measure of network complexity is an important
prerequisite for further studies of evolutionary complexity trends.

\section{Representation Language}

One very simple implementation language for undirected graphs is to
label the nodes $1..N$, and the links by the pair $(i,j), i<j$ of
nodes that the links connect. The linklist can be represented simply
by a $N(N-1)/2$ length bitstring, where the $\frac12j(j-1)+i$th
position is 1 if link $(i,j)$ is present, and 0 otherwise. We also
need to prepend the string with the value of $N$ in order to make it
prefix-free --- the simplest approach is to interpret the number of
leading 1s as the number $N$, which adds a term $N+1$ to the measured
complexity.

\def\noder{.1}

\begin{table}
\psset{unit=4mm}
\begin{center}
\begin{tabular}{c|l}
Network & Bitstring description \\\hline
\pspicture(2,2)
\cnode(1,1.73){\noder}{A}
\cnode(0,0){\noder}{B}
\cnode(2,0){\noder}{C}
\ncline{A}{B}
\endpspicture & 1110100, 1110010, 1110001\\
\pspicture(2,2)
\cnode(1,1.73){\noder}{A}
\cnode(0,0){\noder}{B}
\cnode(2,0){\noder}{C}
\ncline{A}{B}
\ncline{A}{C}
\endpspicture & 1110110, 1110101, 1110011\\
\pspicture(-0.5,-1)(0.5,0.86)
\cnode(0,0){\noder}{A}
\cnode(-0.86,0.5){\noder}{B}
\cnode(0.86,0.5){\noder}{C}
\cnode(0,-1){\noder}{D}
\ncline AB
\ncline AC
\ncline AD
\endpspicture & 11110110100, 11110101010, 11110011001, 11110000111\\
\end{tabular}
\end{center}
\caption{A few example networks, with their bitstring descriptions}
\label{net-examp}
\end{table}

Some example 3 and 4 node networks are shown in table
\ref{net-examp}. One can see how several descriptions correspond to
the same topological network, but with different node numberings.

A few other properties are also apparent. A network $A$ that has a
link wherever $B$ doesn't, and vice-versa might be called a
complement of $B$. A bitstring for $A$ can be found by inverting the
1s and 0s in the linklist part of the network description. Obviously,
$\omega(A,L)=\omega(B,L)$.

The empty network, and the fully connected network have linklists that
are all 0s or 1s. These networks are maximally complex at
\begin{equation}\label{emptyC}
{\cal C}=\frac12N(N+1)+1
\end{equation}
bits. This, perhaps surprising feature, is partly a consequence of the
definition we're using for network equivalence. If instead we ignored
unconnected nodes (say we had an infinite number of nodes, but a only
a finite number of them connected into a network), then the empty
network would have extremely low complexity, as one would need to sum
up the $\omega$s for $N=0,1,\ldots$. But in this case, there would no
longer be any symmetry between a network and its complement.

It is also a consequence of not using a Turing complete representation
language. Empty and full networks are highly compressible, therefore
we'd expect a Turing complete representation language would be able to
represent the network in a compressed form, lowering the measured
complexity.

\def\three{
\cnode(0,0.7){\noder}{A}
\cnode(-.6,-.5){\noder}{B}
\cnode(.6,-.5){\noder}{C}
}

\begin{table}
\psset{unit=4mm}
\begin{center}
\begin{tabular}{c|c|r|r}
Network & Complement & $\omega$ & ${\cal C}$ \\\hline
\pspicture(-1,-1)(1,1)
\three 
\endpspicture
&
\pspicture(-1,-1)(1,1)
\three 
\ncline AB
\ncline AC
\ncline BC
\endpspicture & 1 & 7 \\
\pspicture(-1,-1)(1,1)
\three 
\ncline AB
\endpspicture
&
\pspicture(-1,-1)(1,1)
\three 
\ncline AC
\ncline BC
\endpspicture & 3 & 5.42 \\\hline
\end{tabular}
\end{center}
\caption{Enumeration of all 3-node networks, with number of equivalent
  bitstrings ($\omega$) and complexity (${\cal C}$)}
\label{3net-table}
\end{table}

\def\four{
\cnode(-.7,-.7){\noder}{A}
\cnode(-.7,.7){\noder}{B}
\cnode(.7,-.7){\noder}{C}
\cnode(.7,.7){\noder}{D}
}

\begin{table}
\psset{unit=4mm}
\begin{center}
\begin{tabular}{c|c|r|r}
Network & Complement & $\omega$ & ${\cal C}$ \\\hline
\pspicture(-1,-1)(1,1)
\four
\endpspicture
&
\pspicture(-1,-1)(1,1)
\four
\ncline AB
\ncline AC
\ncline AD
\ncline BC
\ncline BD
\ncline CD
\endpspicture & 1 & 11 \\
\pspicture(-1,-1)(1,1)
\four
\ncline BD
\endpspicture
&
\pspicture(-1,-1)(1,1)
\four
\ncline AB
\ncline AC
\ncline AD
\ncline BC
\ncline CD
\endpspicture & 6 & 8.42 \\
\pspicture(-1,-1)(1,1)
\four
\ncline BA
\ncline BD
\endpspicture
&
\pspicture(-1,-1)(1,1)
\four
\ncline AC
\ncline AD
\ncline BC
\ncline CD
\endpspicture & 12 & 7.42 \\
\pspicture(-1,-1)(1,1)
\four
\ncline CA
\ncline BD
\endpspicture
&
\pspicture(-1,-1)(1,1)
\four
\ncline AB
\ncline AD
\ncline BC
\ncline CD
\endpspicture & 3 & 9.42 \\
\pspicture(-1,-1)(1,1)
\four
\ncline CA
\ncline BD
\ncline BA
\endpspicture
&
same & 12 & 7.42 \\
\pspicture(-1,-1)(1,1)
\four
\ncline DA
\ncline BD
\ncline BA
\endpspicture
&
\pspicture(-1,-1)(1,1)
\four
\ncline AC
\ncline BC
\ncline CD
\endpspicture & 4 & 9 \\\hline
\end{tabular}
\end{center}
\caption{Enumeration of all 4-node networks, with number of equivalent
  bitstrings ($\omega$) and complexity (${\cal C}$)}
\label{4net-table}
\end{table}

Networks of 3 nodes and 4 nodes are sufficiently simple that it is
possible enumerate all possibilities by hand. It is possible to
numerically enumerate larger networks using a computer, however one
will rapidly run into diminishing returns, as the number of bitstrings
to consider grows as $2^{\frac12N(N-1)}$. I have done this up to 8
nodes, as shown in Fig. \ref{cl8}. 

\section{Computing $\omega$}

The first problem to be solved is how to determine if two network
descriptions in fact correspond to the same network.  We borrow a
trick from the field of symbolic computing, which is to say we arrange
a canonical labeling of the nodes, and then compare the canonical
forms of each description. Brendan McKay \citeyear(1981){McKay81} has
solved the problem of finding canonical labelings of arbitrary
graphs, and supplies a convenient software library called {\em
  nauty}\footnote{Available from http://cs.anu.edu.au/\~{}bdm/nauty.}
that implements the algorithm. 

The number of possible distinct descriptions is given by $N!$ (the
number of possible renumberings of the nodes), {\em divided} by the
number of such renumberings that reproduce the canonical
form. As a stroke of good fortune, nauty reports this value as the
order of the automorphism group, and is quite capable of computing
this value for networks with 10s of thousands of nodes within seconds
on modern CPUs. So the complexity value ${\cal C}$ in equation
(\ref{complexity}) is computationally feasible, with this particular
choice of representation.

%
%
%
%
%

\section{Compressed complexity and Offdiagonal complexity}

I have already mentioned the issue of non Turing completeness of the
proposed bitstring representation of a network. This has its most
profound effect for regular networks, such as the empty or full
networks, where ${\cal C}$ is at a maximum, yet contained a great deal
of redundancy in the expression. To get a handle on how much
difference this might make, we can try a compression algorithm of the
all the equivalent bitstring representations, choosing the length most
compressed representation as a new measure I call zcomplexity.
Inspired by the brilliant use of standard compression programs (gzip,
bzip2, Winzip etc.) to classify texts written in an unknown
language\cite{Benedetto-etal02}, I initially thought to use one of
these compression libraries. However, all of the usually open source
compression libraries were optimised for compressing large computer
files, and typically had around 100 bits of overhead. Since the
complexities of all networks studied here are less than around 50
bits, this overhead precludes the use of standard techniques.

So I developed my own compression routine, based around run length
encoding, one of the simplest compression techniques. The encoding is
simple to explain: Firstly a ``wordsize'' $w$ is chosen such that
$\log_2 N \leq w \leq \log_2 N + \log_2 (N-1) -1$. Then the
representation consists of $w$ 1 bits, followed by a zero, then $w$
bits encoding $N$, then the compressed sequence of links. Repeat
sequences are represented by a pair of $w$ bit words, which give the
repeat count and length of a sequence, followed by the sequence to be
repeated. As an example, the network:
\begin{displaymath}
1111110101010101010101
\end{displaymath}
can be compressed to
\begin{displaymath}
\underbrace{111}_{w}0\underbrace{110}_{N}\underbrace{000}_{\mathrm{rpt}}
\underbrace{010}_{\mathrm{len}}\underbrace{10}_{\mathrm{seq}}.
\end{displaymath}
Here 000 represents 8, not 0, as a zero repeat count makes no sense!
Also, since the original representation is prefix free, the extra 0
that the compressed sequence adds to the original is ignored.

By analogy with equation (\ref{complexity}) define zcomplexity as 
\begin{equation}\label{zcomplexity}
{\cal C}_z = 1+\log_2\sum_b 2^{-\min\{ \zeta(b), N(N+1)/2+1\}}\\
\end{equation}
where $b$ iterates over all bitstring representations of the network we're
measuring, and $\zeta(b)$ is the compressed length of $b$, using the
best $w$, by the aforementioned compression algorithm. The extra 1
takes into account a bit used to indicate whether the compressed or
uncompressed sequence is used, so ${\cal C}_z\leq{\cal C}+1$.

The optimal $w$ for the empty (or full) network
$w=\lceil\log_2N\rceil$, and zcomplexity can be readily
computed as
\begin{eqnarray}
  {\cal C}_z &=& 2+3w+\lceil\frac{N(N-1)}{2^{w+1}}\rceil \nonumber\\
  &=& \left\{
\begin{array}{ccc}
  2+3\log_2 N + \frac{N-1}2 & \mathrm{if}& n=2^w\\
  5+3\log_2(N-s) + \lceil\frac{N(N-1)}{4(N-s)}\rceil & \mathrm{if} & n=2^{w-1}+s\\
\end{array}
\right.
\end{eqnarray}
Compared with equation (\ref{emptyC}), we can see that it makes a
substantial difference. Already at $N=5$, ${\cal C}_z=13$ and ${\cal
  C}=16$, with the difference increasing with $N$.

To compute the zcomplexity for an arbitrary graph, we need to iterate
over all possible bit representations of a graph. There are two
obvious ways to do this, since the number of nodes $N$, and number of
links $l$ are identical in all representations:
\begin{itemize}
\item Start with the bitstring with the initial $l$ bits of the
  linkfield set to 1, and the remaining bits 0. Then iterate over all
  permutations, summing the right hand term into a bin indexed by the
  canonical representation of the network. This algorithm computes
  ${\cal C}_z$ for all networks of $N$ nodes and $l$ links. This
  algorithm has complexity
  $\left(\begin{array}{c}N(N-1)/2\\l\end{array}\right) = N(N-1)\ldots(N-l)/l!$
\item Take the network, and iterate over all permutations of the node
  labels. Some of these permutations will have identical bitstring
  representations as others --- as each bitstring is found, store it
  in a set to avoid double counting. This algorithm has complexity $N!$
\end{itemize}
In my experiments, I calculate zcomplexity for all networks with
linkcount $l$ such that
$\left(\begin{array}{c}N(N-1)/2\\l\end{array}\right) < N!$, then
sample randomly networks with greater link counts. 

\begin{figure}
\begin{center}
\epsfbox{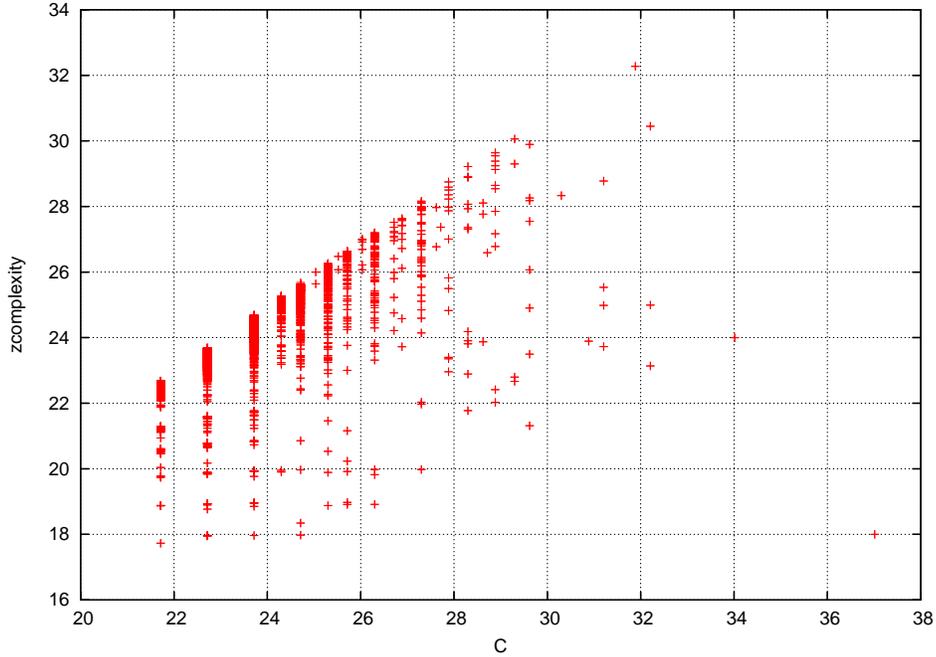}
\end{center}
\caption{${\cal C}_z$ plotted against ${\cal C}$ for all
networks of order 8. Note the empty/full network lying in the lower
right hand corner}
\label{cl8}
\end{figure}

Fig. \ref{cl8} shows ${\cal C}_z$ plotted against ${\cal C}$ for all
networks of order 8, which is about the largest size network for which
an exhaustive computation of ${\cal C}_z$ is feasible.

Unfortunately, without a smarter way of being able to iterate over
equivalent bitstring representations, zcomplexity is not a feasible
measure, even it more accurately represents complexity. The disparity
between ${\cal C}_z$ and ${\cal C}$ is greatest for highly structured
graphs, so it would be interest to know when we can use ${\cal C}$,
and when a more detailed calculation is needed.

Claussen\citeyear(2004){Claussen04} introduced a measure he calls {\em
  offdiagonal complexity}, which measures the entropy of the
distribution of links between different node degree. Regular graphs
will have zero offdiagonal complexity, as the node degree distribution
is sharply peaked, and takes on moderate values for random graphs
(where node degree distribution is roughly exponential) and is
extremal for scale-free graphs. Since the discrepancy between ${\cal
  C}$ and ${\cal C}_z$ was most pronounced with regular graphs, I
looked at offdiagonal complexity as a predictor for this
  discrepancy. 
  
\begin{figure}
\begin{center}
\epsfbox{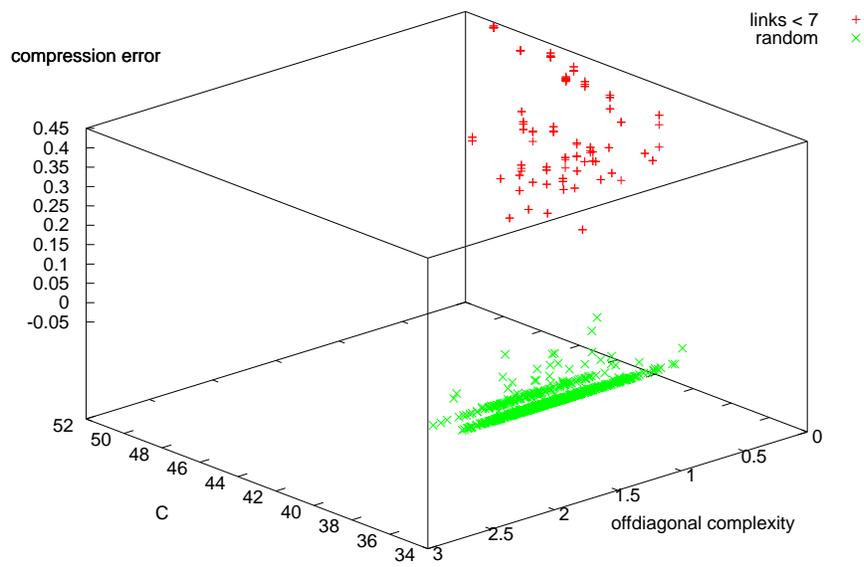}
\end{center}
\caption{Compression error as a function of ${\cal C}$ and offdiagonal
  complexity for networks with 10 nodes. All networks with link count
  less than 7 were evaluated by method 1, and 740 graphs with more
  than 7 links were selected at random, and computed using method 2.
  The separation between the two groups is due to compressibility of
  sparse networks.}
\label{cl10}
\end{figure}

Figure \ref{cl10} shows the compression error (defined as $\frac{{\cal
    C}-{\cal C}_z}{{\cal C}}$) plotted as a function of offdiagonal
complexity and ${\cal C}$. The dataset falls clearly into two groups
--- all sparse networks with link count less than 7, and those graphs
sampled randomly, corresponding to the two different methods mentioned
above. The sparse networks are expected to be fairly regular, hence
have high compression error, whereas randomly selected networks are
most likely to be incompressible, hence have low compression error.

\begin{figure}
\begin{center}
\epsfbox{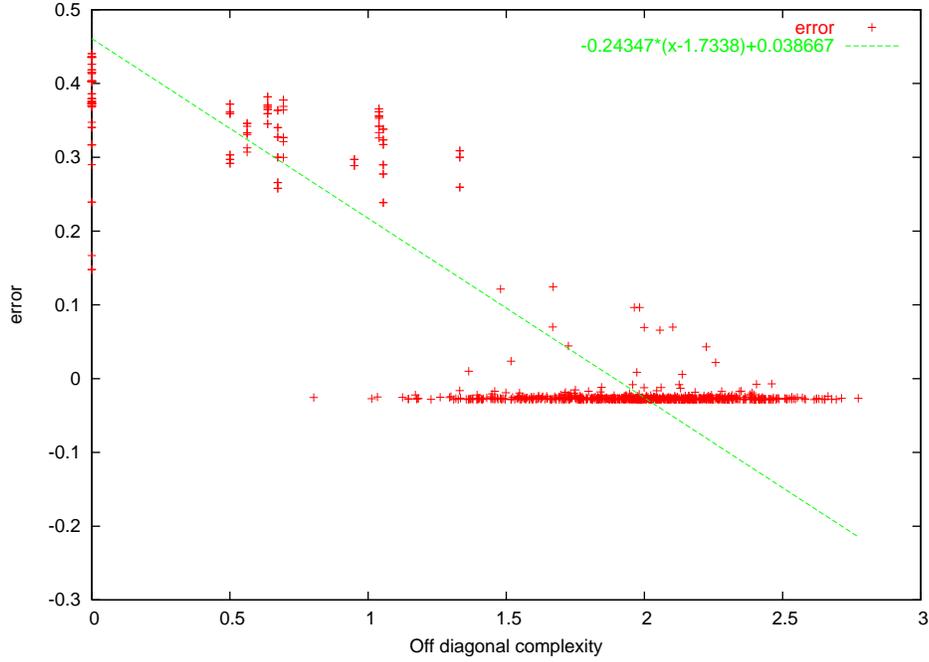}
\end{center}
\caption{Compression error as a function of offdiagonal complexity. A
  least squares linear fit is also shown}
\label{odc-error}
\end{figure}

Figure \ref{odc-error} shows the results of a linear regression
analysis on offdiagonal complexity with compression error. The
correlation coefficient is -0.87. So clearly offdiagonal complexity is
correlated (negatively) with compression error, much as we expected,
however it is not apparently a good test for indicating if the
compression error is large. A better distinguishing characteristic is
if ${\cal C}$ is greater than the mean random ${\cal C}$ (which can be
feasibly calculated) by about 3--4 bits. What remains to be done is to
look at networks generated by a dynamical process, for example
Erd\"os-R\'enyi random graphs\cite{Erdos-Renyi59}, or
Barab\'asi-Albert preferential attachment\cite{Barabasi-Albert99} to
see if they fill in the gap between regular and algorithmically random
graphs.

\section{Conclusion}

In this paper, a simple representation language for $N$-node
undirected graphs is given. An algorithm is presented for computing
the complexity of such graphs, and the difference between this
measure, and one based on a Turing complete representation language is
estimated. For most graphs, the measure presented here computes
complexity correctly, only graphs with a great deal of regularity are
overestimated. 

A code implementing this algorithm is implemented in C++ library, and
is available from version 4.D17 onwards as part of the \EcoLab{}
system, an open source modelling framework hosted at
http://ecolab.sourceforge.net.

Obviously, undirected graphs is simply the start of this work --- it
can be readily generalised to directed graphs, and labeled graphs
such as food webs (although if the edges a labeled by a real value,
some form of discretisation of labels would be needed).

Furthermore, most interest is in complexities of networks generated
by dynamical processes, particularly evolutionary processes. Some of
the first processes that should be examined are the classic
Erd\"os-R\'enyi random graphs and Barab\'asi-Albert preferential attachment.

\section*{Acknowledgements}

I wish to thank the {\em Australian Centre for Advanced Computing and
  Communications} for a grant of computer time, without which this
  study would not be possible.

\bibliographystyle{plain}
\bibliography{rus}

\end{document}